\documentclass[12pt,preprint]{aastex}

\slugcomment{To appear in ApJ}

\shorttitle{What Causes P-mode Asymmetry Reversal?}
\shortauthors{D. Georgobiani, R. F. Stein, {\AA}. Nordlund}

\begin{document}

\title{What Causes P-mode Asymmetry Reversal? \\
}

\author{Dali Georgobiani and Robert F. Stein}
\affil{Department of Physics and Astronomy,
Michigan State University, 1312 Biomedical Physical Sciences,
East Lansing, MI, USA 48824-2320}
\email{dali@pa.msu.edu, steinr@pa.msu.edu}
\and
\author{{\AA}ke Nordlund}
\affil{Teoretisk Astrofysik Center, Denmark
Grundforskningsfond, Juliane Maries Vej 30, DK-2100
K{\o}benhavn \O, Denmark}
\email{aake@astro.ku.dk}

\begin{abstract}
The solar acoustic p-mode line profiles are asymmetric.
Velocity spectra have more power on the low-frequency
sides, whereas intensity profiles show the opposite sense of
asymmetry.
Numerical simulations of the upper convection zone have resonant
p-modes with
the same asymmetries and asymmetry reversal as the observed modes.
The temperature and velocity power spectra at
optical depth $\tau_{\rm cont} = 1$ have the opposite asymmetry as
is observed for the intensity and velocity spectra.
At a fixed geometrical depth, corresponding to $<\tau_{\rm cont}>=1$,
however, the temperature and velocity spectra have the same asymmetry.
This indicates that the asymmetry reversal is produced by
radiative transfer effects and not by correlated noise.
\end{abstract}


\section{Introduction}

For almost a decade, it has been known that the power spectra
of solar acoustic modes are asymmetric, velocity has more power on the
low frequency side and intensity has more power on the high frequency
side of the power maxima \citep[e.g.][]{duv93}.
The asymmetry
reversal between velocity and intensity is thought to be due to the
correlated background noise contribution to the intensity power
spectra (Nigam et al. 1998). It is debated whether
the asymmetry reversal occurs in velocity \citep{rav97}
or intensity \citep{nig98} or both \citep{kab99}.
Theoretical models predict asymmetries that depend on the source depth
and type \citep{kab99,dg00b}. \citet{rav97}
considered the superposition of dipole and quadrupole sources;
\citet{nig98} used a combination of monopole and dipole
terms; \citet{kab99} show that the asymmetry reversal
could be triggered even by dipole or quadrupole sources alone.

Simulations of the shallow upper layer of the solar convective zone
have resonant acoustic modes like the Sun.  The emergent intensity
and the velocity in the photosphere have asymmetric spectra with the
opposite asymmetry \citep{dg00a}. In this letter we
calculate the
temperature and velocity power spectra at the continuum optical depth
$\tau = 1$ and at the geometrical depth corresponding to $<\tau>=1$.
At unit continuum optical depth the velocity and temperature
have opposite asymmetry, with the velocity having more low-frequency
power and the intensity more high frequency power.  At fixed
geometrical depth, however, the velocity and temperature have the
same asymmetry, more low frequency power, for the fundamental mode.
These results
indicate that the asymmetry reversal is caused by
radiative transfer effects, and not by correlated noise.

\section{Model of the Solar Convection}

We use a three - dimensional hydrodynamic code of 
\citet[and references therein]{san98} to simulate the upper layers of the
solar convection zone. The computational domain covers 6 Mm by 6 Mm
horizontally 
and 3 Mm in vertical direction, from 0.5 Mm above the $\tau = 1$ surface
to 2.5 Mm beneath it. The model includes non-gray, LTE radiative transfer.
Horizontal boundaries are periodic,
while vertical ones are transmitting. The spatial resolution is
100 km horizontally and $\sim$ 50 km vertically, with a finer grid
interpolated for solving the radiative transfer equation.  The
radiation field is calculated by solving the Feautrier equation
along a vertical and 4
straight inclined rays, after averaging the Planck function into four
bins by wavelength sorted according to opacity 
\citep[cf][]{nor82,nas90,nas91}. Snapshots are saved
at 30 s intervals.  We have simulated 43 hours of solar time.

\section{Calculation of Power Spectra and Phase Relations}

We compare power spectra of intensity, temperature
and velocity in order to shed light on a cause of the p-mode line
profile asymmetry and its reversal between velocity and
intensity. From our simulations, we have coordinate- and
time-dependent quantities: vertical velocity $V(x,y,z,t)$, temperature
$T(x,y,z,t)$, emergent intensity $I(x,y,t)$, etc.
To obtain a power spectrum at a
particular geometrical depth $z_0$, we use $V$ and $T$ at
$z = z_0$.  To obtain $V$ and $T$ at a particular
optical depth, say, $\tau = 1$, we calculate $\tau(x,y,z,t)$ and
interpolate V and T to the height at which $\tau(x,y,z,t)=1$ for each
position $(x,y)$ and time $t$.  We
choose $z_0$ to be the depth for which $<\tau> = 1$.  Clearly, $V(x,y)$
or $T(x,y)$ at a
geometrical depth $z_0$ will not be the same as $V(x,y)$ or
$T(x,y)$ at $\tau = 1$:  we see deeper in the cooler intergranular
lanes compared to the hotter granules.  We investigate
if this makes a difference in the power spectra and somehow affects
line asymmetries.

We separate the oscillation modes into radial modes, for which we
average our data horizontally, and
nonradial modes, for which we multiply the data by the corresponding spatial
sines or cosines and then average horizontally 
\citep[for more details, see][]{dg00a}.
We then Fourier - transform these
time strings to get power spectra and phase relations. In our simulations,
the first nonradial mode (horizontal wavelength $\lambda_h$ = 6 Mm)
corresponds to
a harmonic degree $\ell=740$, because the simulation box represents
a small fraction of the solar surface.  We investigate the
behavior of the emergent intensity, plasma temperature and plasma
velocity for this first non-radial mode of the simulation
\citep[see also][]{dg00a}.

\section{Results}

\begin{figure}
\plotone{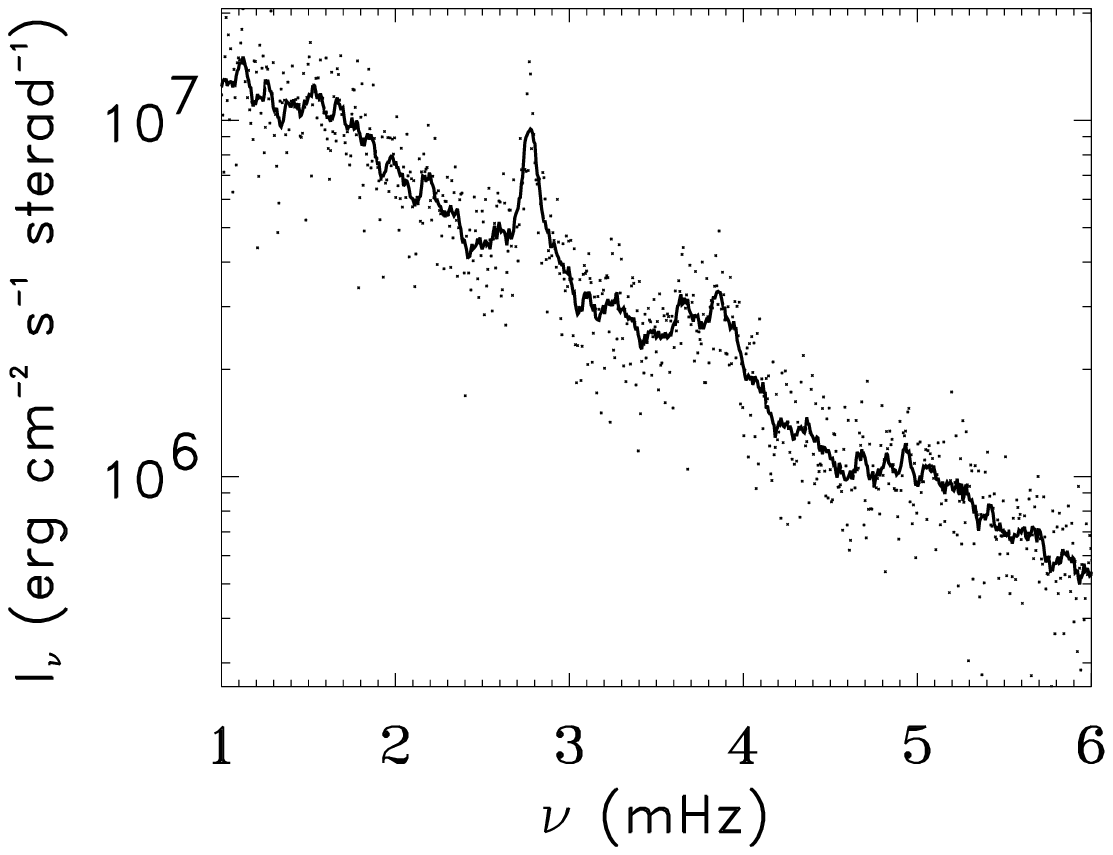}
\caption{Spectrum of intensity, $I_{\nu}$.
The solid curve is boxcar smoothed over $\Delta \nu = 65 \mu$Hz.
\label{intensity}}
\end{figure}

\begin{figure}
\plotone{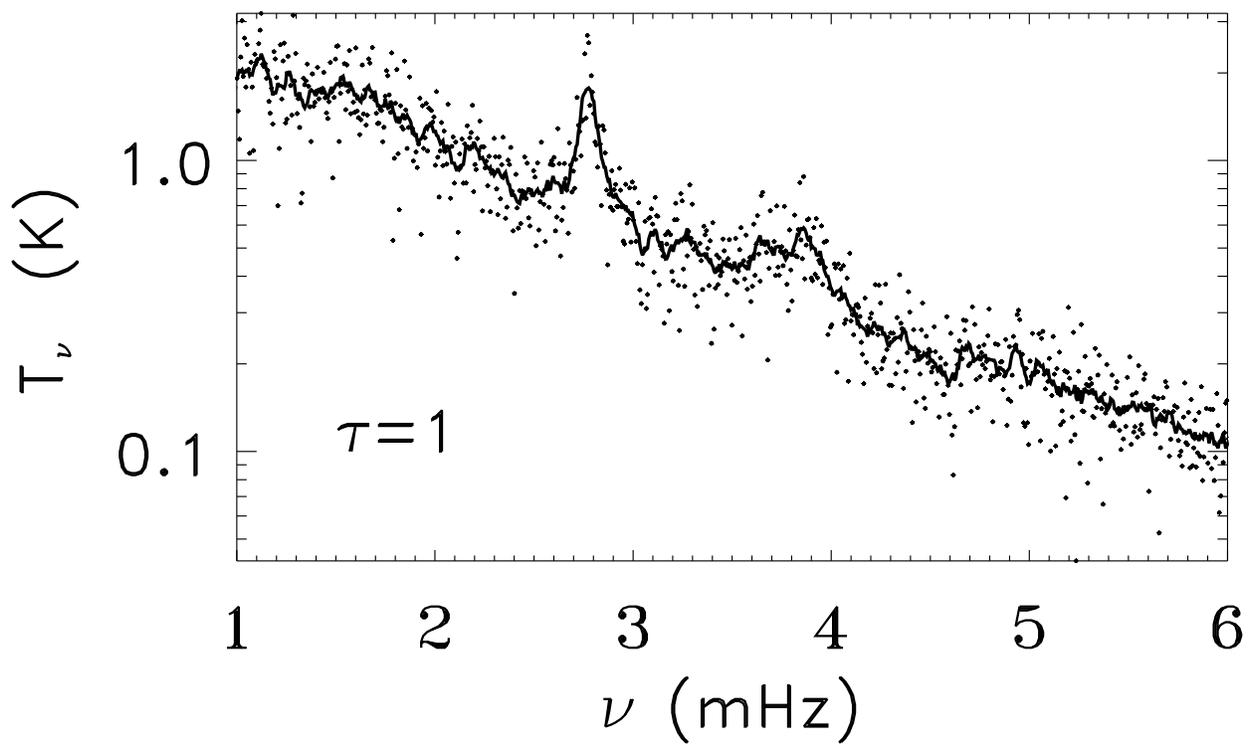}
\caption{Spectrum of temperature, $T_{\nu}$, at $\tau(x,y,t) = 1$.
Again, the solid curve is smoothed. \label{T-tau}}
\end{figure}

\begin{figure}
\plotone{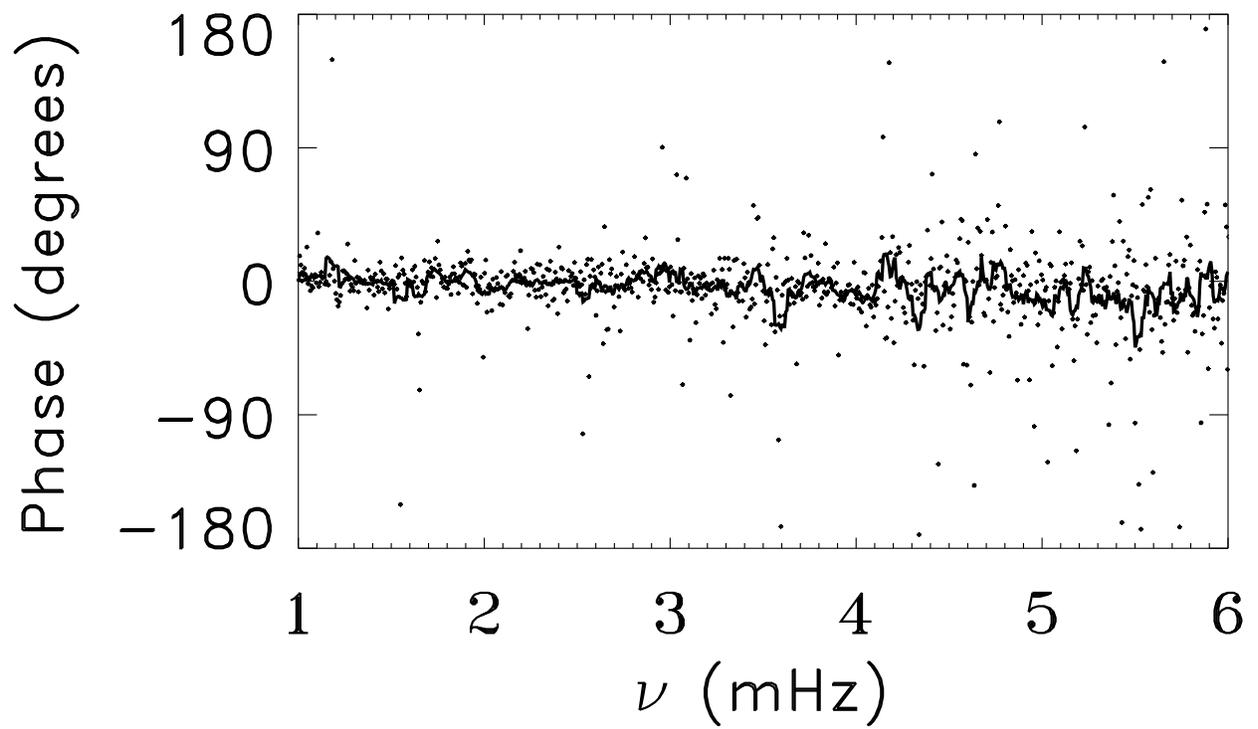}
\caption{Phase difference between the intensity and the temperature
measured at the $\tau = 1$ level. \label{I-Tphase}}
\end{figure}

\begin{figure}
\plotone{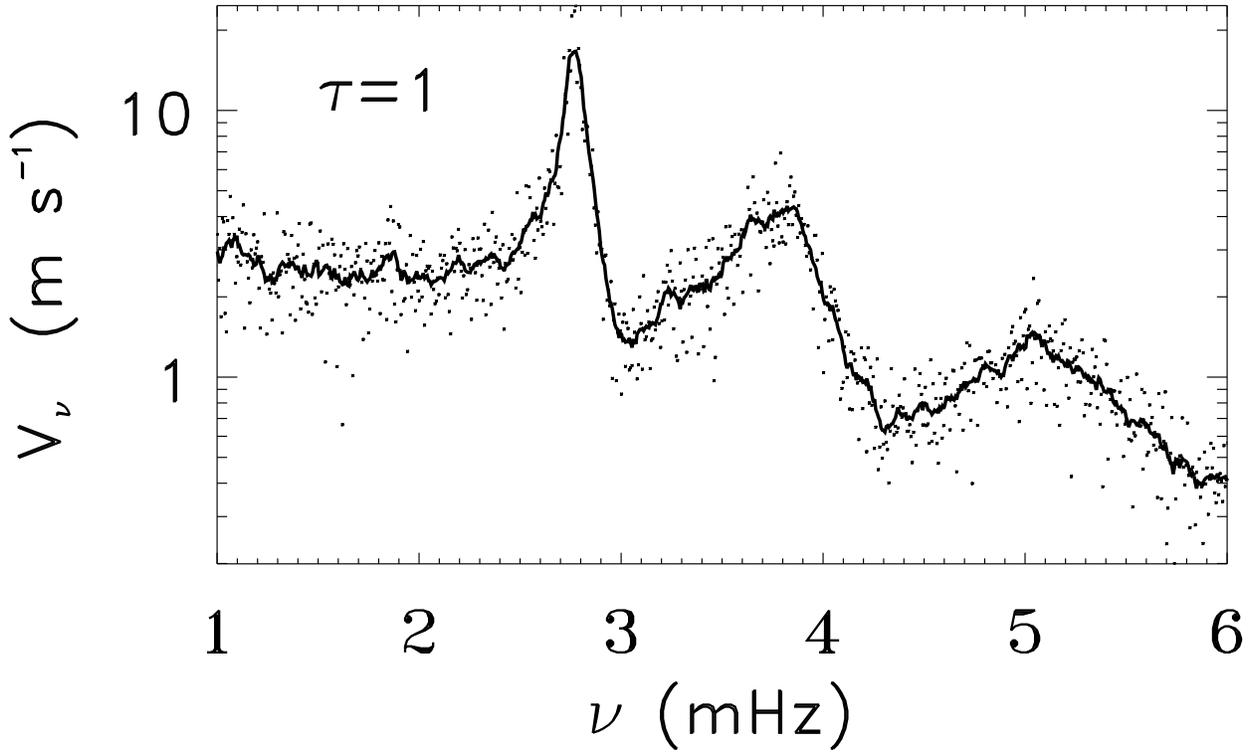}
\caption{Spectrum of velocity, $V_{\nu}$, at $\tau (x,y,t)= 1$.  
Its spectrum is almost the same at fixed geometrical depth
corresponding to $<\tau>_{x,y,t}=1$. \label{V-tau}}
\end{figure}

\begin{figure}
\plotone{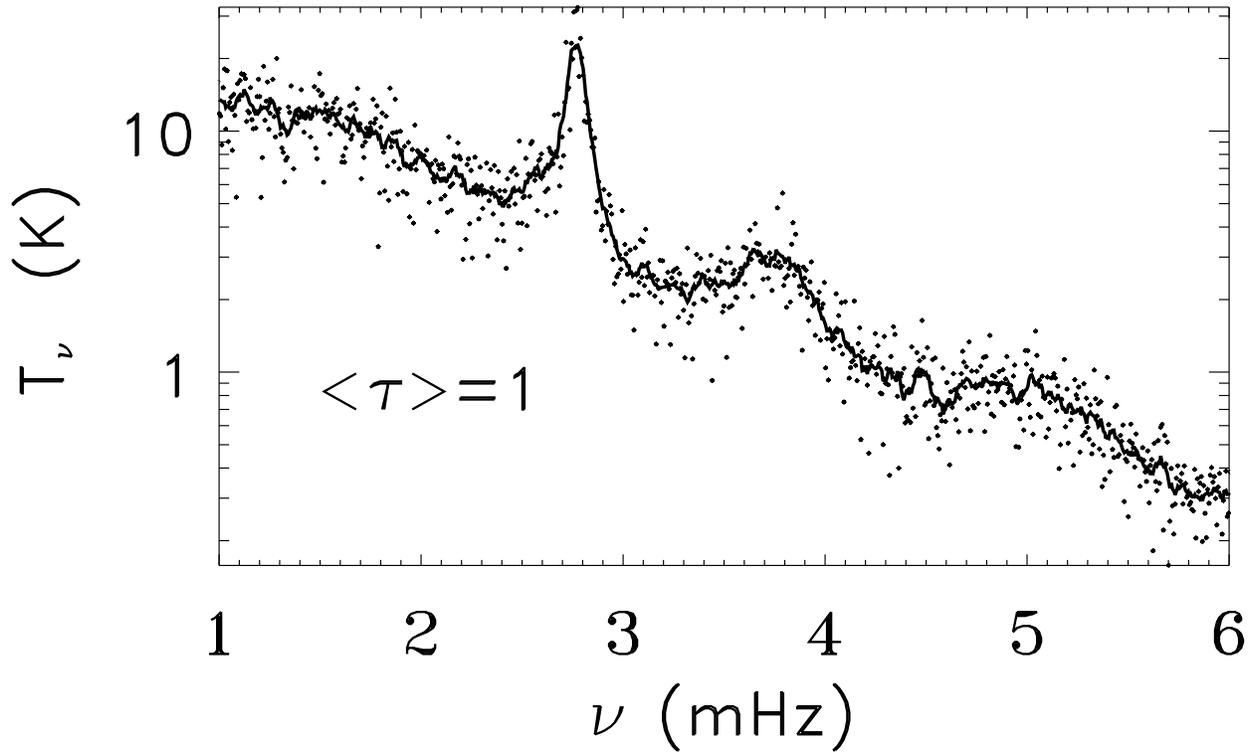}
\caption{Spectrum of temperature , $T_{\nu}$,  at $<\tau> = 1$.
The temperature at $<\tau> = 1$ has the same asymmetry as the 
velocity. \label{T-z}}
\end{figure}

The emergent continuum intensity (Fig.~\ref{intensity}) is a good
measure of the plasma temperature at local instantaneous continuum
optical depth unity, $\tau(x,y,t)=1$, as expected from the
Eddington-Barbier relations.  The temperature at $\tau=1$ has the
same spectrum (Fig.~\ref{T-tau}) and phase (Fig.~\ref{I-Tphase}) as
the emergent intensity.  These phase differences are essentially zero
for all frequencies.
The spectra of the emergent intensity and plasma temperature at
$\tau=1$ have opposite asymmetry to the plasma velocity
(Fig.~\ref{V-tau}) as is observed \citep[e.g.][]{duv93}
and as was discussed by \citet{dg00a}.
The spectrum of the temperature at fixed geometrical depth $z_0$,
corresponding to the average continuum optical depth unity,
$<\tau>_{x,y,t} = 1$, (Fig.~\ref{T-z}) is, however, rather different
from its spectrum at local instantaneous unit optical depth
(Fig.~\ref{T-tau}), with different asymmetries especially noticeable
for the fundamental mode.  The temperature at $<\tau>=1$ has the
same asymmetry as the velocity.

\section{Discussion}

\begin{figure}
\plotone{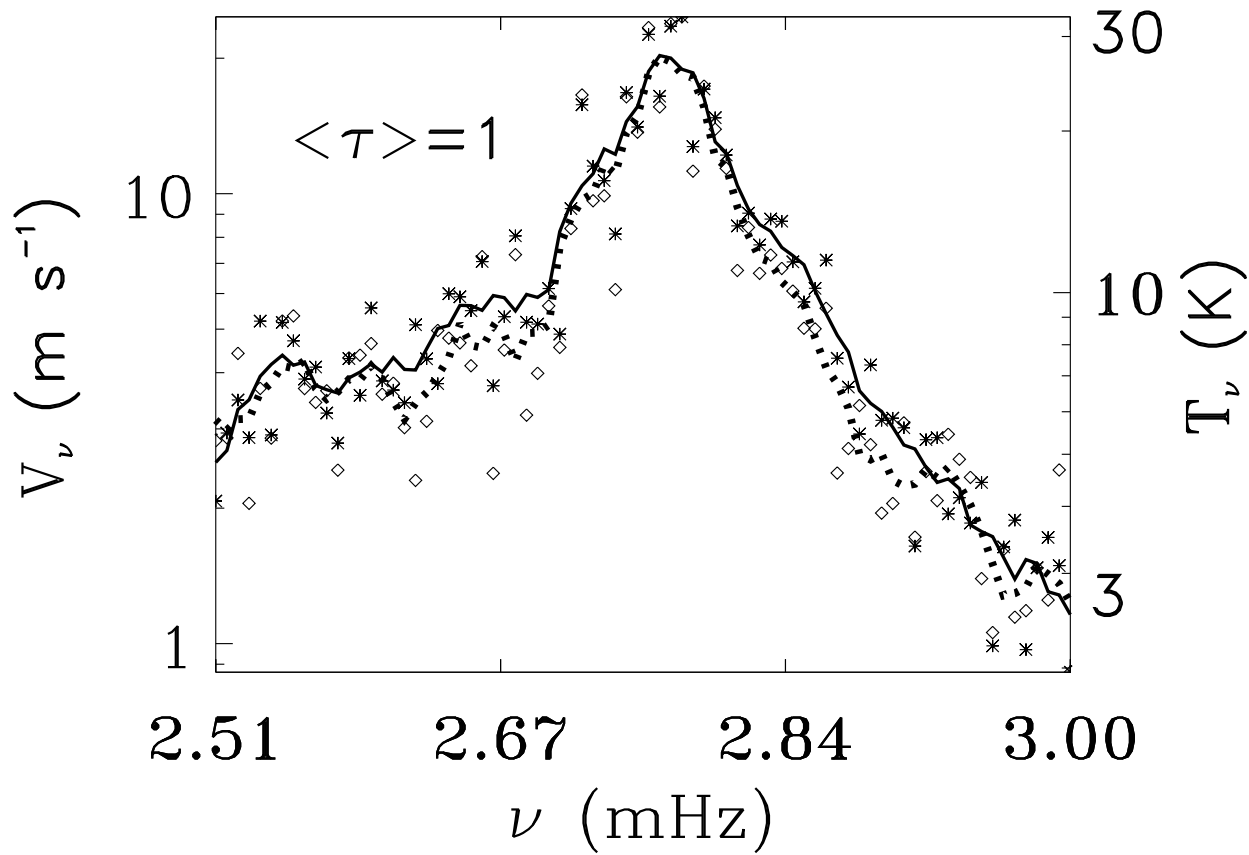}
\caption{The temperature (dotted) and velocity (solid) spectrum for the
first non-radial fundamental mode measured at $<\tau> = 1$.
Here and after, diamonds represent unsmoothed temperature profiles,
whereas stars correspond to the parameter on the left y-axis.
The temperature and velocity profiles look very similar to each other.
Curves are smoothed over 32 $\mu$Hz. \label{V-Tavgtau}}
\end{figure}

\begin{figure}
\plotone{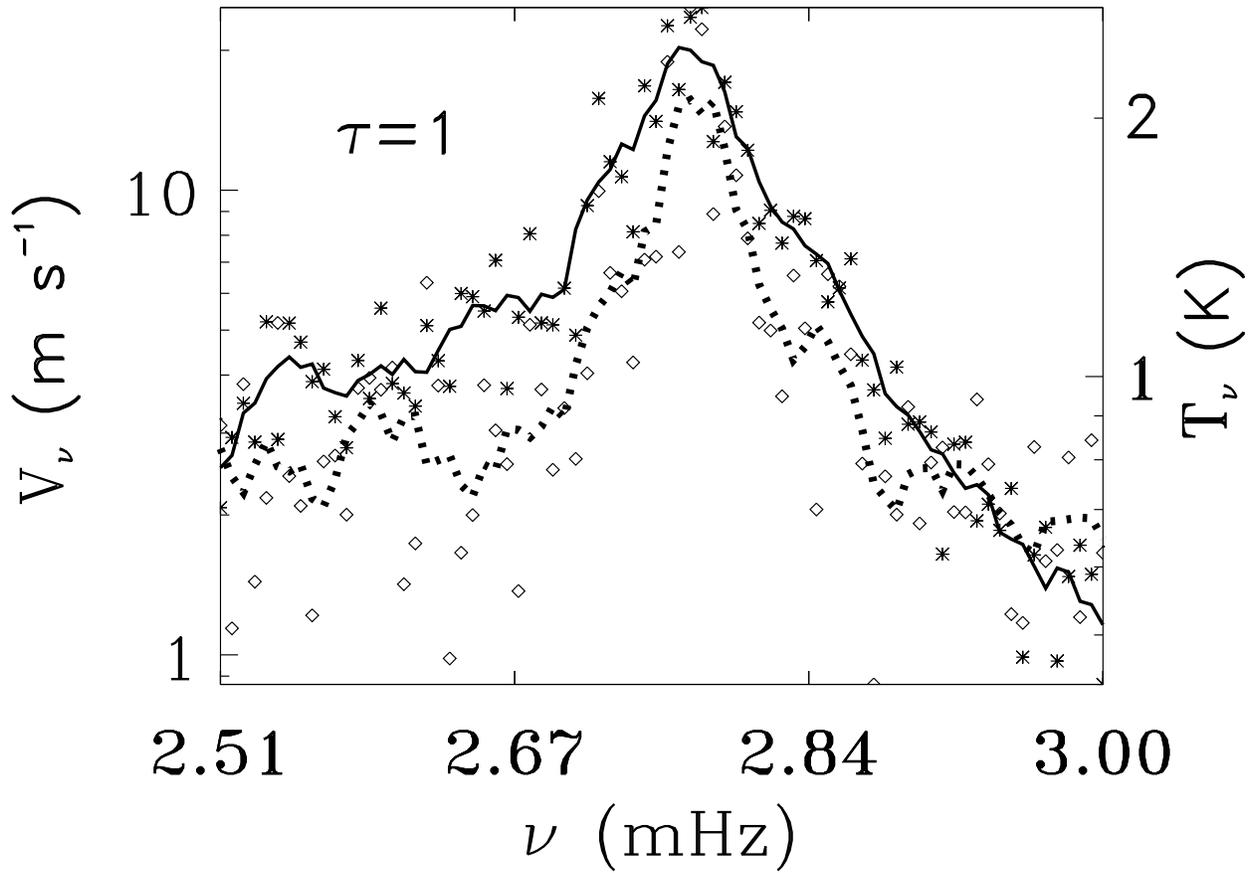}
\caption{The temperature (dotted) and velocity (solid) spectrum for the
first non-radial fundamental mode measured at $\tau = 1$.  The
amplitude of the temperature fluctuations is nonuniformly reduced
across the mode peak. \label{V-Ttau}}
\end{figure}

What changes the asymmetry of the temperature spectrum between
measuring it at local $\tau = 1$ and average $<\tau> = 1$?
We analyze the first non-radial fundamental mode that has the most
prominent asymmetry and agrees closely with the corresponding
solar $\ell=740$ mode.  Figure \ref{V-Tavgtau} shows the velocity
and temperature profile of this mode at average continuum optical
depth one.  It is clear that the velocity and temperature have the
same profiles.  Figure \ref{V-Ttau} shows the velocity and
temperature profiles measured at local optical depth one, which is
where one would see them.  The velocity spectrum is hardly changed,
but the amplitude of
the temperature fluctuations is reduced by more than an order of
magnitude.  The high temperature sensitivity of the $H^{-}$
opacity obscures high temperature gas and alters the height at which
the gas is observed.  This reduces the magnitude of the observed
temperature fluctuations, but this reduction is not as great on the
high frequency side as on the low frequency side.  Hence, the mode
asymmetry is changed. 

\begin{figure}
\plotone{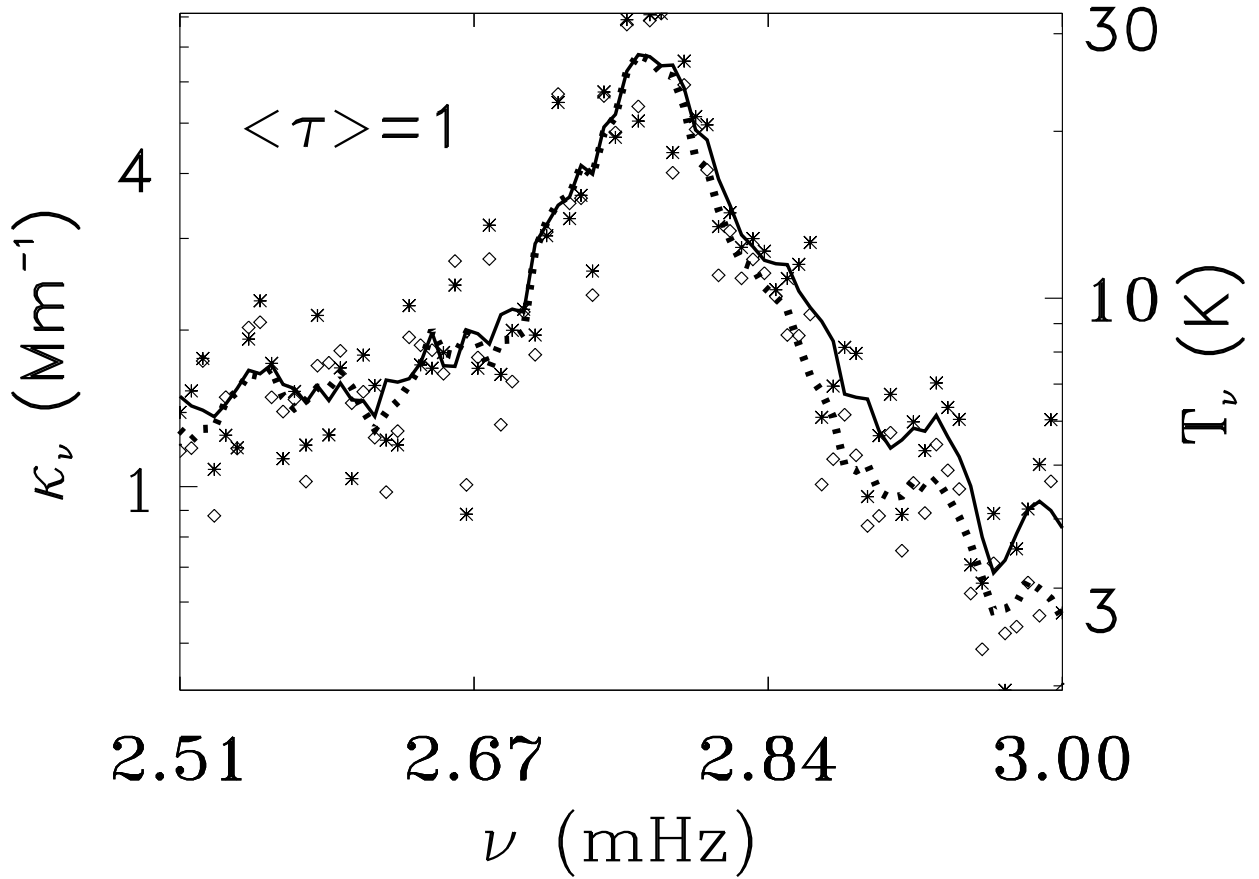}
\caption{The temperature (dotted) and opacity (solid) spectrum for the
first non-radial fundamental mode measured at $<\tau> = 1$.  The
larger temperature fluctuations on the low frequency side of the mode
produce larger opacity variations. \label{K-Tavgtau}}
\end{figure}

\begin{figure}
\plotone{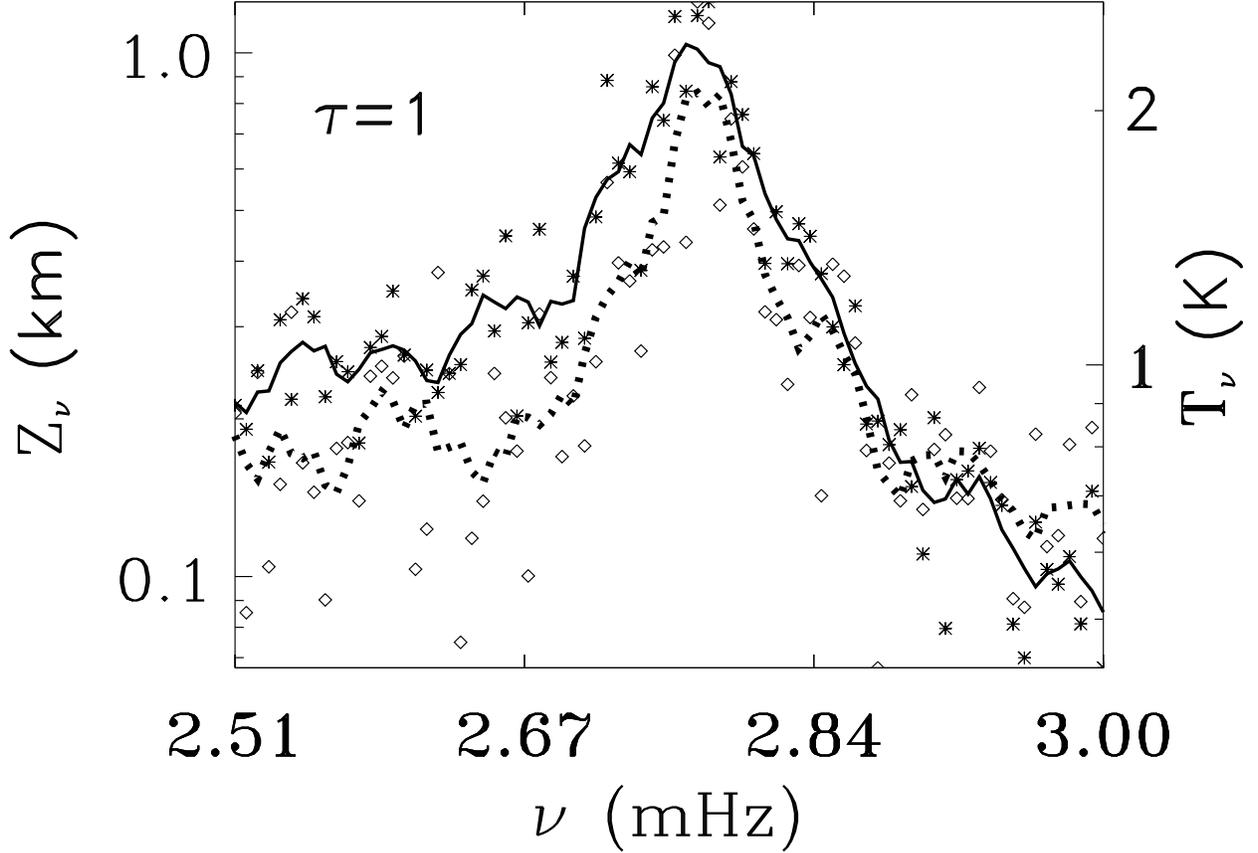}
\caption{The spectrum of the height of local optical depth one,
$z_{\tau(x,y,t)=1}$, (solid) and temperature (dotted) measured at
local $\tau = 1$.  The location of local optical depth unity varies
more on the low frequency side of the mode, where the opacity
variation is larger due to the larger temperature fluctuations.  This
reduces the temperature fluctuations at local optical depth unity
more on the low frequency side of the mode compared to the high
frequency side and, in accordance with the Eddington-Barbier relations,
leads to smaller intensity fluctuations on the low frequency side of
the mode and larger intensity fluctuations on the high frequency side
of the mode. \label{Z-Ttau}}
\end{figure}

\begin{figure}
\plotone{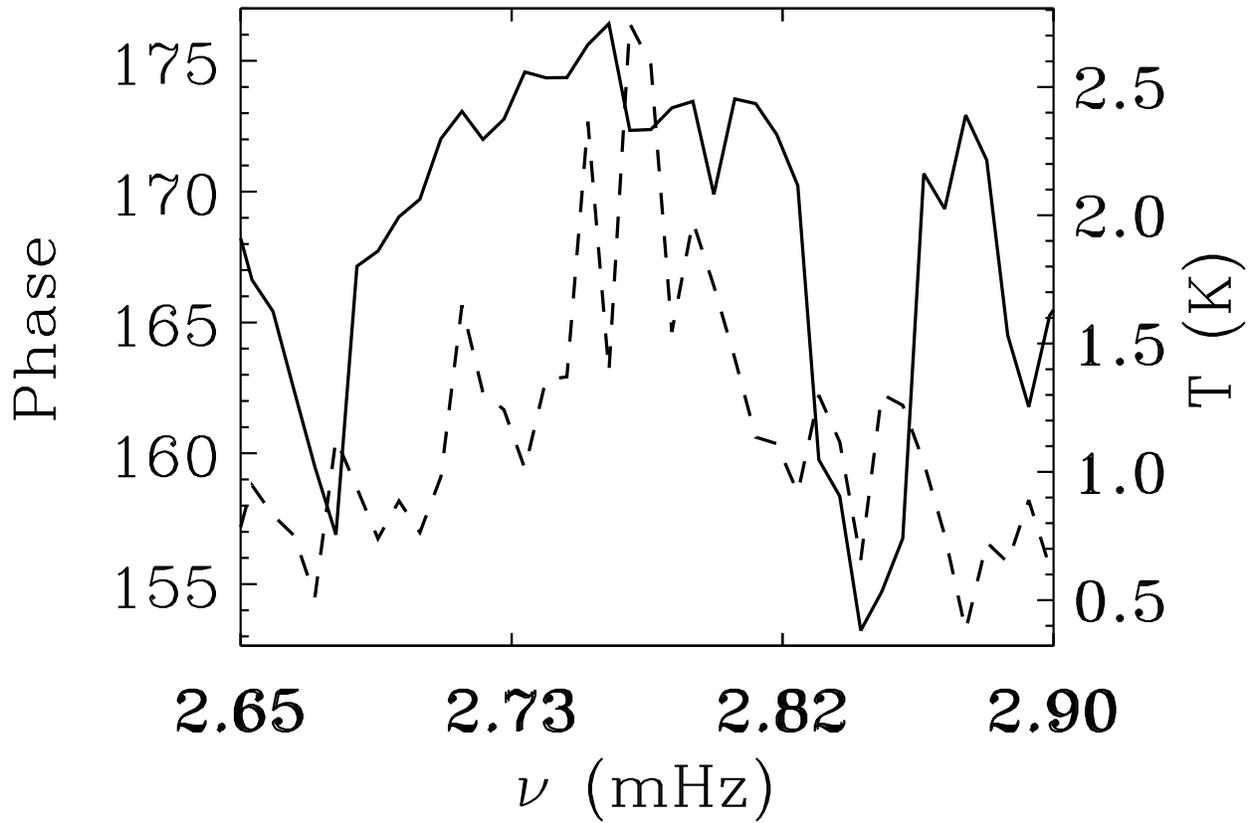}
\caption{The phase (solid line) between the height of $\tau = 1$ and the
temperature at $<\tau> = 1$.  Phase $\approx 180^o$ means that the height
of unit optical depth is greatest (smallest z and lowest temperature)
when the temperature at $<\tau>= 1$ is largest.  Also shown is
temperature at $\tau =1$ (dashed). \label{Z-T_phase}}
\end{figure}

Why
is the reduction of the temperature fluctuations different at high and
low frequencies?  At the fixed geometrical height $<\tau>=1$, the
temperature fluctuations were larger on the low frequency side of the
mode.  This produces a larger opacity variation on the low frequency
side of the mode (Fig.~\ref{K-Tavgtau}), which in turn leads to a larger
variation in the height where local $\tau(x,y,t)=1$
(Fig.~\ref{Z-Ttau}).  The radiation temperature we see is
equal to the gas temperature at optical depth unity,
according to the Eddington-Barbier relations.
The phases of temperature and height of unit
optical depth are such that where the temperature at fixed geometrical
depth is largest we observe the temperature at greatest height (smallest
z) (Fig~\ref{Z-T_phase}).  Since the gas
temperature is decreasing outward, the larger variation in the location
where the radiation originates on the low frequency side of the mode
leads to a smaller temperature variation there, while the smaller
variation in the location where radiation originates on the high
frequency side of the mode leads to a larger temperature variation
there.  We therefore observe a larger
intensity variation on the high frequency side of the mode than on
the low frequency side of the mode and thus the asymmetry is reversed
from that observed in the velocity (and temperature when measured at a
fixed geometrical depth).

\section{Conclusion}

We have found that the emergent intensity and the temperature spectra
at local instantaneous optical depth unity have the {\em opposite}
asymmetry to the velocity as is observed, while the temperature
at fixed geometrical depth corresponding to average optical depth
unity has the {\em same} asymmetry as the velocity.  This indicates that
radiation transfer plays a crucial role in the asymmetry reversal observed
between the intensity and Doppler velocity, and that this reversal is
not due solely to effects of correlated noise.  The mode asymmetry
has been shown to depend on the separation of observation location
and source location for the case of a simple localized
$\delta$-function source \citep{kab99,dg00b}.
T($\tau=1$) and T($<\tau>=1$) are observed at different heights.
Oscillation induced opacity changes vary the location of radiation
emission ($\tau=1)$ in a way that reduces the magnitude of the
temperature fluctuations and reverses their asymmetry.

\acknowledgments
This work was supported in part by NASA grants NAG 5 9563 and NAG 5 8053,
NSF grant AST 9819799, and by a grant of computer time from the
National Center for Supercomputing Applications (which is supported
by NSF).  Their support is greatly appreciated and was crucial to the
completion of this work.

\clearpage

\clearpage

\clearpage

\clearpage 

\clearpage

\clearpage

\clearpage

\clearpage

\clearpage

\clearpage

\clearpage

\end{document}